\documentclass[preprint2]{aastex}

\newcommand{\ba}{\begin{eqnarray}}
\newcommand{\ea}{\end{eqnarray}}
\newcommand{\be}{\begin{equation}}
\newcommand{\ee}{\end{equation}}
\newcommand{\Par}{\parallel}
\newcommand{\Perp}{\perp}

\slugcomment{to be submitted to \apj}

\shorttitle{Electron Heating in Hot Accretion Flows}
\shortauthors{Sharma et al.}

\begin{document}

\title{Electron Heating in Hot Accretion Flows}

\author{Prateek Sharma, Eliot Quataert}
\affil{Astronomy Department, University of California,
    Berkeley, CA 94720}
\email{psharma@astro.berkeley.edu, eliot@astro.berkeley.edu}

\author{Gregory W. Hammett}
\affil{Princeton Plasma Physics Laboratory, Princeton, NJ 08543}
\email{hammett@pppl.gov}

\and

\author{James M. Stone}
\affil{Department of Astrophysical Sciences, Princeton University, Princeton, NJ 08544}
\email{jstone@astro.princeton.edu}

\begin{abstract}

Local (shearing box) simulations of the nonlinear evolution of the
magnetorotational instability in a collisionless plasma show that
angular momentum transport by pressure anisotropy ($p_\Perp \ne
p_\Par$, where the directions are defined with respect to the local
magnetic field) is comparable to that due to the Maxwell and Reynolds
stresses.  Pressure anisotropy, which is effectively a large-scale
viscosity, arises because of adiabatic invariants related to $p_\Perp$
and $p_\Par$ in a fluctuating magnetic field.  In a collisionless
plasma, the magnitude of the pressure anisotropy, and thus the
viscosity, is determined by kinetic instabilities at the cyclotron
frequency.  Our simulations show that $\sim 50$ \% of the
gravitational potential energy is directly converted into heat at
large scales by the viscous stress (the remaining energy is lost to
grid-scale numerical dissipation of kinetic and magnetic energy).  We
show that electrons receive a significant fraction ($\sim~[T_e/T_i]^{1/2}$)
of this dissipated energy.  Employing this heating
by an anisotropic viscous stress in one dimensional models of
radiatively inefficient accretion flows, we find that the radiative
efficiency of the flow is greater than 0.5\% for $\dot{M} \gtrsim
10^{-4} \dot{M}_{Edd}$. Thus a low accretion rate, rather than just a
low radiative efficiency, is necessary to explain the low luminosity
of many accreting black holes.  For Sgr A* in the Galactic Center, our
predicted radiative efficiencies imply an accretion rate of $\approx 3
\times 10^{-8} M_\odot \, {\rm yr^{-1}}$ and an electron temperature
of $\approx 3 \times 10^{10}$ K at $\approx 10$ Schwarzschild radii; the
latter is consistent with the brightness temperature inferred from
VLBI observations.

\end{abstract}

\keywords{accretion, accretion disks -- MHD -- plasmas -- Galaxy:  center}

\section{Introduction}

Local and global simulations have shown that the magnetorotational
instability (MRI) is the likely source of turbulence and angular
momentum transport in many astrophysical accretion disks
\citep[e.g.,][]{bal91,bal98,haw95,arm98}.  The MRI has been studied in
a wide variety of contexts, from the disk of the Milky Way
\citep[][]{pio05} to planet-forming disks around young stars (e.g., Winters, 
Balbus, \& Hawley 2003).
Our recent work has focused on the evolution of the MRI in
collisionless plasmas (\citealt{qua02}, Sharma, Hammett, \& Quataert 2003,
\citealt{sha06} [hereafter SHQS]; for related work, see
\citealt{bal04} and \citealt{kro06}).  These calculations are
motivated by the application to radiatively inefficient accretion flow
(RIAF) models \citep{ich77,ree82,nar95}, which are believed to be
applicable to accretion onto black holes and neutron stars when the
accretion rate is less than a few \% of Eddington.  Sgr A* in our
Galactic Center is the paradigm case for accretion via a RIAF (see
\citealt{qua03} for a review).

One of the central problems in the theory of hot accretion flows
remains understanding the extreme low luminosity of accretion onto
black holes in many galactic nuclei (\citealt{fab88,fab95,loe01}) and
X-ray binaries (e.g., Narayan, McClintock, \& Yi 1996). In the context of RIAF models,
the observed low luminosities can be explained if most of the
available mass is not accreted because of outflows or convection
(Blandford \& Begelman 1999; Quataert \& Gruzinov 2000; Stone, Pringle, \& 
Begelman 1999; Narayan, Igumenshchev, \& Abramowicz 2000), or if the gravitational potential
energy primarily heats the poorly radiating ions
\citep[][]{ree82,nar95}.  Global MHD simulations support the former
possibility \citep{sto01,haw01}.

A number of analytic arguments have been presented for the heating of
electrons and ions in the collisionless plasmas appropriate to RIAFs.
These estimates focus on heating by MHD turbulence
\citep{qua98,gru98,black99,qg99,med00} or reconnection
\citep{bis97,qg99}. Our shearing box simulations of MRI turbulence in
a collisionless plasma suggest an additional mechanism for particle
heating in RIAFs.  Adiabatic invariance ($\mu=v_\perp^2/B \propto
p_\perp/B = $ constant) in a collisionless plasma with a fluctuating
magnetic field results in a macroscopic pressure anisotropy, with the
pressure perpendicular to the field lines ($p_\Perp$) exceeding that
along the field lines ($p_\Par$).  This pressure anisotropy results in
a viscous transport of angular momentum that is comparable to the
transport by magnetic stresses (SHQS).  In a collisional plasma, the
magnitude of pressure anisotropy, and thus the viscosity, is
determined by Coulomb collisions.  In turn, the relative heating of
electrons and ions by viscosity is set by the temperature and
particle-mass dependence of the Coulomb cross-section (which results
in primarily ion heating).  In a collisionless plasma, however, the
pressure anisotropy---and thus the viscosity---is regulated by the
growth of small scale instabilities that violate adiabatic invariance
and isotropize the plasma pressure (e.g., the firehose, mirror, ion
cyclotron, and electron whistler instabilities).  As we shall show,
these instabilities also regulate the viscous heating of electrons and
ions in a collisionless accretion flow.

In \S 2 we summarize the equations that describe the evolution of the
pressure tensor ($p_\Par$ and $p_\Perp$); a more detailed presentation
is given in SHQS.  We then present upper limits on the pressure
anisotropy because of gyroradius scale instabilities.  We use these
limits to provide analytic estimates for the viscous heating of
electrons and ions in a collisionless plasma and contrast these
results with the more familiar results appropriate to a collisional
plasma.  In \S 3 we discuss the results of two-component (electron +
proton) shearing-box simulations of the MRI.  This is an extension of
our single component simulations (SHQS).  In \S 4 we use 1-D models of
RIAFs to calculate the radiative efficiency using the viscous heating
of electrons calculated in \S 2 and \S 3. We also show that electron
Coulomb collisions are less effective at reducing the pressure
anisotropy in RIAFs than kinetic microinstabilities and thus that the
kinetic approach focused on throughout this paper is appropriate.
Finally, \S 5 is a discussion of our key results and their application
to RIAFs.

\section{Analytic considerations}

In the limit that the fluctuations of interest have length scales
much larger than the gyroradius and time scales much longer than the
cyclotron period, the equations describing the evolution of $p_\Par$
and $p_\Perp$ are given by \citep[][and references
therein]{kul83,sny97} 
\ba
\label{eq:SB4}
\nonumber
\frac{\partial p_{\Par,s}}{\partial t} &+&
\nabla \cdot (p_{\Par,s} {\bf V}) + \nabla \cdot {\bf q}_{\Par,s} + 2 p_{\Par,s}
{\bf \hat{b}\hat{b}} : \nabla {\bf V} \\
&-& 2q_{\Perp,s}
\nabla \cdot {\bf \hat{b}} = -\frac{2}{3} \nu_{eff,s} (p_{\Par,s} - p_{\Perp,s}), \\
\label{eq:SB5}
\nonumber \frac{\partial p_{\Perp,s}}{\partial t} &+& \nabla \cdot
(p_{\Perp,s} {\bf V}) + \nabla \cdot {\bf q}_{\Perp,s} + p_{\Perp,s}
\nabla \cdot {\bf V} \\ \nonumber &-& p_{\Perp,s} {\bf \hat{b}\hat{b}}
: \nabla {\bf V} + q_{\Perp,s} \nabla \cdot {\bf \hat{b}} \\ &=&
-\frac{1}{3} \nu_{eff,s} (p_{\Perp,s} - p_{\Par,s}), \ea
where the subscript `$s$' denotes the species (electrons and
ions [we only consider protons]), ${\bf \hat{b}}$ is the local
magnetic field direction, ${\bf V}$ is the fluid velocity, ${\bf
q_{\Perp}} = q_\Perp {\bf \hat{b}} \, ({\bf q_{\Par}} = q_{\Par}
{\bf \hat{b}})$ is the flux of perpendicular (parallel) thermal
energy along the local magnetic field.  The effective pitch-angle
scattering rate due to high frequency waves (or Coulomb collisions)
is given by $\nu_{eff}$.

Equations (\ref{eq:SB4}) and (\ref{eq:SB5}) can be obtained by taking
the moments of the drift kinetic equation, the Vlasov equation in the
limit of large wavelength and long timescales.  These moment equations
run into the usual ``closure" problem.  In our numerical calculations,
we use a simple closure for the heat flux, in which the heat flux is
proportional to the temperature gradient, with the conductivity being
a free parameter.  This model can roughly reproduce kinetic effects
such as collisionless damping of linear modes
\citep[see][]{sha03,sha06b}.  For the analytic estimates in this
section, these details of the heat fluxes are not important. In
addition, our numerical simulations show that the electron heating
rate does not depend strongly on the details of the heat flux model.

Equations (\ref{eq:SB4}) \& (\ref{eq:SB5}) can be combined to derive
equations for the internal energy ($e_s \equiv 3p_s/2 \equiv
p_{\Par,s}/2+p_{\Perp,s} $) and pressure anisotropy ($\Delta p_s =
p_{\Par,s}-p_{\Perp,s}$), which are given by 
\ba
\label{eq:internal_energy}
\nonumber \frac{\partial e_s}{\partial t} &+& {\bf \nabla \cdot} \left
( e_s {\bf V + q_s}\right ) + p_{\Perp,s} {\bf \nabla \cdot V} \\
&=& - \Delta p_s {\bf \hat{b}\hat{b} : \nabla V}, \\
\label{eq:pressure_anisotropy}
\nonumber \frac{\partial \Delta p_s}{\partial t} &+& {\bf \nabla \cdot
} \left ( \Delta p_s {\bf V} + {\bf \Delta q_s} \right ) = -3 p_s {\bf
\hat{b}\hat{b} : \nabla V} \\ \nonumber &+& p_{\Perp,s} {\bf \nabla
\cdot V} + 3 q_{\Perp,s} {\bf \nabla \cdot \hat{b}} \\ &-& \nu_{eff,s}
\Delta p_s, \ea
where ${\bf q_s}={\bf q_{\Par,s}}/2
+ {\bf q_{\Perp,s}}$ and ${\bf \Delta q_s} = {\bf q_{\Par,s} -
q_{\Perp,s}}$. In statistically steady MRI-driven turbulence,
$\Delta p_s < 0$ on average (see, e.g., Fig. 4 of SHQS). This sign
of the pressure anisotropy corresponds to the expected outward
transport of angular momentum by the viscous stress.

If the fluctuations of interest are relatively incompressible then the
volume averaged version of equation (\ref{eq:internal_energy}) is
given by \be
\label{eq:pressure_evolution}
\frac{3}{2}\frac{d}{dt}p_s \sim q^+_{V,s} \equiv -\Delta p_s {\bf
\hat{b}\hat{b} : \nabla V}.  \ee The RHS of equation
(\ref{eq:pressure_evolution}) represents heating due to the
anisotropic pressure. In an accretion flow, the velocity gradient in
equation (\ref{eq:pressure_evolution}) can be roughly decomposed into
two components, that due to the background shear $\Omega(r)$, and that
due to the turbulent velocity fluctuations $\delta {\bf V}$ (such a
decomposition is formally made in the shearing box simulations
described in \S 3, see e.g., \citealt{haw95}).  Thus the heating rate
itself can be decomposed into two contributions: a heating term given by
\be q^+_{V1,s} \equiv
\left ( d \Omega \over d \ln r \right ) \Delta p_s b_r b_\phi,
\label{eq:visc} \ee
and one given by \be q^+_{V2,s} \equiv - \Delta p_s {\bf \hat{b}\hat{b} :
\nabla \delta V}. \label{eq:fluctuating} \ee Equation (\ref{eq:visc})
represents viscous heating due to the background shear while equation
(\ref{eq:fluctuating}) represents dissipation of the turbulent
fluctuations (by, e.g., collisionless Landau damping).\footnote{In our
numerical simulations, there is also energy lost due to grid-scale
viscosity and resistivity; these will be discussed in \S3.}

In the analytic estimates that follow we focus on viscous heating due
to the background shear; in our simulations, we find that this is the
dominant contribution to the total heating in equation
(\ref{eq:pressure_evolution}). Note that, just as for a collisional
viscosity, this term represents the direct conversion of ordered
kinetic energy (differential rotation) into heat at large
scales.\footnote{Collisionless viscous heating is analogous to the
betatron mechanism (also known as magnetic pumping) where adiabatic
invariance and pressure isotropizing collisions can result in net
plasma heating \citep[][]{bud61,kul05}.}  The other sources of heating
are less direct: energy is converted into magnetic and velocity
fluctuations which are dissipated by collisionless damping at large
scales, or at small scales via a nonlinear cascade.

We now consider the collisional (Braginskii) and collisionless limits
of equation (\ref{eq:pressure_evolution}) and quantify the viscous
heating in each of these limits.

\subsection{The Collisional limit}

If the collisional mean free path is small compared to the gradient
length scales of interest then the
viscous stress in a plasma can be written as a pressure tensor,
anisotropic with respect to the field lines \citep{bra65}, where the
magnitude of the pressure anisotropy is limited by collisional pitch
angle scattering (see, e.g., \citealt{sny97} for the equivalence of
Braginskii viscosity and anisotropic pressure). Pressure anisotropy in
the collisional limit of equation (\ref{eq:pressure_anisotropy}) is given
by $\Delta p_s = -(p_s/\nu_s) 3 {\bf
\hat{b}\hat{b}:\nabla V}$, where $\nu_s$ is the (Coulomb) collisional
pitch angle scattering rate. Using $\nu_s \propto m_s^{-1/2}
T_s^{-3/2}$ equation (\ref{eq:pressure_evolution}) becomes \be
\label{eq:heating_Braginskii}
q_{V,s}^+ \propto m_s^{1/2} \, T_s^{5/2} \, \left({\bf
\hat{b}\hat{b}:\nabla V}\right)^2, \ee where $m_s$ and $T_s$ are the mass and
temperature of the species, respectively. As noted above, for an accretion
flow ${\bf \hat{b}\hat{b}:\nabla V} \sim d\Omega/d\ln r$.  Equation
(\ref{eq:heating_Braginskii}) implies that for $T_e = T_i$, the
heavier ions are heated preferentially by viscosity by a factor of
$\sim 40$.  This conclusion is strengthened if $T_e < T_i$, as is
typically the case in hot accretion flows.

\subsection{The collisionless limit}

When the mean free path is comparable to or larger than the gradient
length scales of interest, the collisional (Braginskii) description is
no longer valid. In the collisionless regime, the pressure anisotropy
is limited by pitch angle scattering due to cyclotron frequency
fluctuations that violate adiabatic invariance.
A minimum level of
pitch angle scattering is that due to kinetic instabilities generated
by the pressure anisotropy itself. We focus on that possibility here.
In principle, the rate of pitch angle scattering can be much higher if
high frequency fluctuations are efficiently generated by a turbulent
cascade, shocks or reconnection.  Whether this in fact occurs is
difficult to assess and may depend on the specific problem of
interest; theories of incompressible MHD turbulence, however, imply
little power near the cyclotron frequency \citep[e.g.,][]{gol95,how06},
and measurements in the solar wind show
significant pressure anisotropy \citep[e.g.,][]{kas02}.

\subsubsection{Upper Limits on the Pressure Anisotropy}

To quantify the rate of pitch angle scattering in a collisionless
plasma, we first discuss limits on the pressure anisotropy due to
kinetic instabilities.  As discussed in \S1 (and SHQS), pressure
anisotropy is naturally created in MRI turbulent plasmas due to the
fluctuating magnetic field.  For $p_\Perp > p_\Par$, as occurs if the
magnetic field is amplified by the MRI, the relevant instabilities are
the mirror and ion-cyclotron instabilities for the ions
(\citealt{has69,gar94}; also discussed in SHQS), and the electron whistler
instability for the electrons \citep[][]{gar96}. In the
opposite limit ($p_\Par>p_\Perp$), the firehose instability reduces
the pressure anisotropy (\citealt{gar98,li00}; SHQS).

The mirror instability arises when the pressure anisotropy becomes
larger than $p_{\Perp,i}/p_{\Par,i}-1 >1/\beta_{\Perp,i}$. However,
the fastest growing mode occurs at the gyroradius scale (and hence
violates $\mu$ invariance and leads to pitch angle scattering) only
when (SHQS; see also \citealt{has69}) \be
\label{eq:mirror_threshold}
\frac{p_{\Perp,i}}{p_{\Par,i}}-1>\frac{7}{\beta_{\Perp,i}}. \ee
For a pressure anisotropy smaller than this threshold, no significant pitch
angle scattering occurs \citep{mck93}.

A plasma is formally unstable to the ion cyclotron instability for
arbitrarily small values of $T_{\Perp,i}/T_{\Par,i} - 1>0$.  However,
the instability can only grow on a reasonable timescale ($\sim$
rotation period) when the pressure anisotropy exceeds a threshold
which can be written as (e.g., \citealt{gar94}; SHQS) \be
\label{eq:IC_threshold}
\frac{T_{\Perp, i}}{T_{\Par, i}} - 1 >
\frac{S_i}{\beta_{\Par,i}^{\alpha_i}}, \ee where $S_i$ and $\alpha_i$
depend on $\gamma/\Omega_i$, the ratio of the growth rate to the ion
cyclotron frequency.  We have used the publically available WHAMP code
\citep[][]{ron82} to calculate $S_i$ and $\alpha_i$ for a specified value
of $\gamma/\Omega_{i}$.  For RIAFs, $\Omega_i \gg \Omega$, the disk
rotation frequency.  Because pressure anisotropy is generated on the
turnover time of the turbulent fluctuations, which is comparable to
the rotation period of the disk ($\sim \Omega^{-1}$), $\gamma \ll
\Omega_i$ is sufficient to generate significant high frequency
fluctuations which provide pitch angle scattering.  Taking
$\gamma/\Omega_i = 10^{-4}$ as a fiducial estimate of the growth rate
required for significant pitch angle scattering, we find $S_i=0.35$
and $\alpha_i=0.45$ as the threshold for the ion cyclotron
instability.  For $\gamma \ll \Omega_i$ the pressure anisotropy
threshold ($S_i$) depends very weakly (roughly logarithmically) on the
growth rate $\gamma/\Omega_i$ so our results are not sensitive to the
particular choice of $\gamma/\Omega_i$ (see, e.g., \citealt{gar94}).

As noted above, the mirror instability and ion cyclotron instability
both act to isotropize the plasma when $p_\perp > p_\parallel$.  For
accretion disks with $\beta_i \lesssim 100$, the ion cyclotron
instability is generally more important than the mirror
instability. In our analytic estimates we assume that the ion cyclotron
instability dominates, but our simulations include both the mirror and
ion cyclotron thresholds discussed above.

Electrons with an anisotropic pressure ($p_{\Perp,e}>p_{\Par,e}$) are
unstable to the whistler instability, which has a frequency $\sim
\Omega_e$ and thus violates the adiabatic invariance of the electrons,
producing pitch angle scattering. The threshold for the growth of the
whistler instability with $\gamma/\Omega_e = 5 \times 10^{-8}$ is
given by\footnote{For non-relativistic electrons, this choice for the
electron growth rate corresponds to the same value of $\gamma$ as that
used in determining the ion cyclotron threshold for the ions above
($\gamma/\Omega_i = 10^{-4}$).}  \be
\label{eq:whistler_threshold}
\frac{T_{\Perp, e}}{T_{\Par, e}} - 1 >
\frac{S_e}{\beta_{\Par,e}^{\alpha_e}}, \ee with $S_e=0.13$ and
$\alpha=0.55$ (again using the WHAMP code).  As with the ion cyclotron
threshold, the whistler threshold depends only weakly on the choice of
$\gamma/\Omega_e$.

In our simulations, we find that $p_{\Perp,s}>p_{\Par,s}$ holds over
most of the computational domain, as expected because of the outward
transport of angular momentum by the viscous stress.  There are,
however, regions where the pressure anisotropy has the opposite
sign. In this case electron and ion firehose instabilities will limit
the pressure anisotropy to \citep[][]{gar98,li00} \be
\label{eq:firehose_threshold}
1-\frac{p_{\Perp,s}}{p_{\Par,s}} < \frac{2}{\beta_{\Par,s}}.  
\ee

The magnitude of the viscous heating in equation
(\ref{eq:pressure_evolution}) is directly proportional to the pressure
anisotropy $\Delta p_s$, which is $< 0$ on average and saturates at a
value close to the $p_{\Perp,s} > p_{\Par,s}$ thresholds discussed
above (as shown explicitly in Fig. \ref{fig:fig3} discussed in the
next section).  For an analytic estimate, we assume that the ion and
electron pressure anisotropies are bounded by the ion cyclotron and
whistler thresholds, respectively.  We then estimate the viscous
heating of electrons and ions in the collisionless limit using
equations (\ref{eq:pressure_evolution}), (\ref{eq:IC_threshold}), and
(\ref{eq:whistler_threshold}) \be
\label{eq:heating_kinetic}
q_{V,s}^+ \propto S_s \beta_s^{1-\alpha_s}. \ee The ion cyclotron and
whistler thresholds correspond to $S_i/S_e \approx 3$ and $\alpha_i
\approx \alpha_e \approx 0.5$, in which case $q_{V,s}^+ \propto
T_s^{1/2}$.  Viscous heating in a collisionless plasma is thus quite
different from that in a collisional plasma: the electron and ion
heating rates are comparable when $T_e \sim T_i$.  Indeed, in the
absence of radiative cooling, equations (\ref{eq:heating_kinetic}) and
(\ref{eq:pressure_evolution}) imply that the electron to ion
temperature ratio approaches $T_e/T_i \sim (S_e/S_i)^2 \sim 0.1$ at
late times, if viscous heating is the dominant heating mechanism (even
if the ions are initially much hotter than the electrons).

\section{Numerical Simulations}

In our previous work (SHQS) we modified the {\tt ZEUS} MHD code
\citep[][]{sto92a,sto92b} to model the dynamics of the MRI in a
collisionless plasma, by including pressure anisotropy, thermal
conduction along field lines, and subgrid models for pitch angle
scattering by microinstabilities when the pressure anisotropy exceeds
the thresholds (discussed in \S 2.2.1).  Our previous work modeled a
single fluid, effectively the ions since they typically dominate the
pressure if $T_i \gg T_e$.  We now extend this work to include both
ions and electrons.  As in SHQS, we restrict ourselves to local
shearing box simulations.

We solve the mass and momentum conservation equations for the shearing
box in MHD (eqs. [35] and [36] in SHQS), where the pressure is the sum
of electron and ion pressures, each evolved according to equations
(\ref{eq:SB4}) and (\ref{eq:SB5}).  Heat fluxes parallel to the field
lines (${\bf q_\Perp}$ and $\bf q_{\bf \Par}$) are given by equations
(40)-(44) of SHQS applied to electrons and ions. The conductivity
along the field lines is given approximately by $\kappa \sim
v_{th}^2/(v_{th} k_L + \nu_{eff})$ where $v_{th}$ is the thermal
velocity of the species, $\nu_{eff}$ is the pitch angle scattering
rate due the microinstabilities, and $k_L$ is a parameter that
corresponds to the typical wavenumber for Landau damping; this gives
the correct collisionless damping rate for a fluctuation whose
wavenumber is $k_L$ but is only approximate for other modes
\citep{sha06b}. For small $\nu_{eff}$, the conductivity in the heat
flux is $\propto k_L^{-1}$ so that small values of $k_L$ correspond to
rapid heat conduction while $k_L \rightarrow \infty$ is the adiabatic
(CGL) limit with no heat conduction.\footnote{CGL refers to the double
adiabatic model of \citet{che56}.}

In our simulations, the usual shearing box boundary conditions are
used (Hawley, Gammie, \& Balbus 1995). Initial conditions are described 
in SHQS.  The
present simulations use a purely vertical initial field with $\beta =
400$. In all runs we initiate a specified electron and ion temperature
ratio at the end of 5 orbits so that our results are not affected by
the strong channel phase at $\sim 4$ orbits that represents the
initial non-linear saturation of the MRI.  Electron and ion cooling is
neglected.  Unless specified otherwise, the resolution is $N_x \times
N_y \times N_z = 27 \times 59 \times 27$, and the box size $L_x\times
L_y \times L_z = 1 \times 2 \pi \times 1$. The initial parameters
in code units are: ion pressure $p_0=10^{-6}$, rotation rate
$\Omega=10^{-3}$, and initial density $\rho = 1$ (as in SHQS). The
thermal conduction parameter $k_L=0.5/\delta z$ for both electrons and
ions, where $\delta z~(=L_z/N_z)$ is the grid spacing in the vertical
direction.  Note that this value of the conductivity corresponds to
correctly capturing Landau damping for modes with wavelengths of $(4
\pi/N_z) L_z \approx 0.5 L_z$, i.e., for modes whose wavelengths are
approximately half the size of the box.

\begin{figure}
\plotone{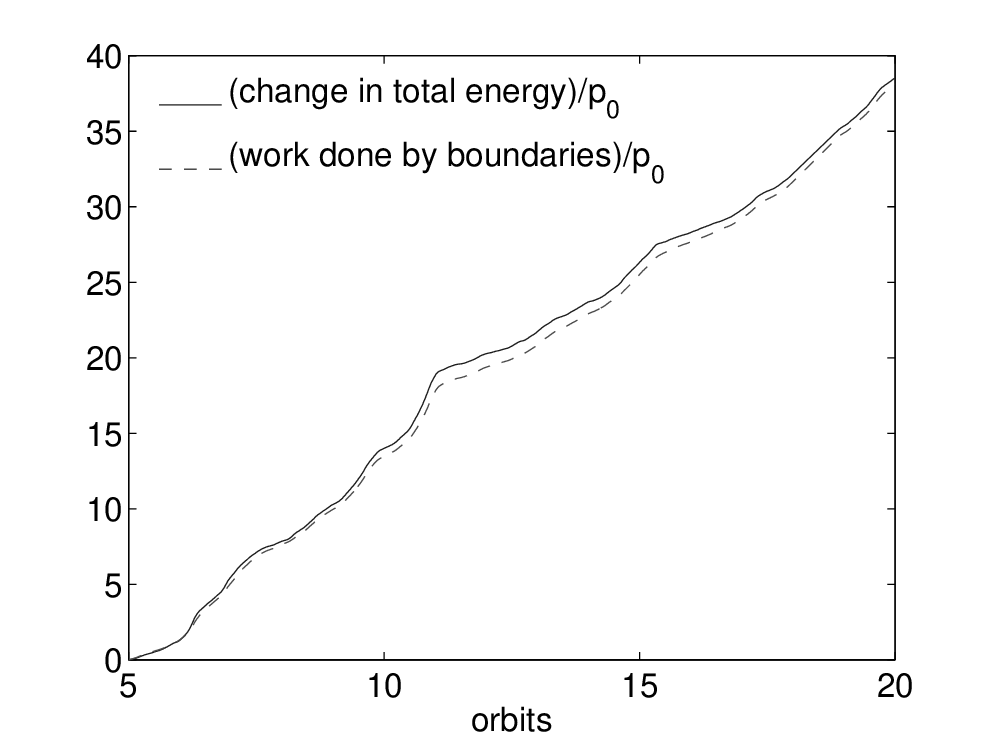}
\caption{The increase in total energy normalized to the initial
pressure $p_0$ (solid line) in shearing box simulations of the MRI,
compared to the work done by the stresses at the boundaries (dashed
line). For a conservative calculation these two curves would agree
identically.  This is not the case
because {\tt ZEUS} is non-conservative.  Nonetheless, our method of
improving energy conservation in {\tt ZEUS} \citep[based on][]{tur03}
conserves energy to better than $\sim 5\%$ over 15 orbits. \label{fig:fig1}}
\end{figure}

Since {\tt ZEUS} is not a conservative code, energy is lost because of
numerical dissipation.  However, to quantify the importance of
different heating mechanisms it is desirable to quantitatively track
the energetics of the accretion flow.  To do so, we have implemented
the method of \cite{tur03} in which energy conservation in {\tt ZEUS}
is significantly improved by keeping track of the kinetic and magnetic
energies lost because of grid-scale averaging. This energy can also,
if desired, be added as a source of plasma heating.  However, we do
not do so because it is physically unclear which species should
receive this energy. In what follows we identify energy lost
numerically in the magnetic field update as ``numerical resistive"
heating ($q^+_{NR}$) and energy lost in the transport step for
momentum as ``numerical viscous" ($q^+_{NV}$) heating.  Compressive
heating, $-p_\Perp {\bf \nabla \cdot V}$, is negligible because the
MRI is relatively incompressible.  In addition to numerical heating
due to grid-scale dissipation, there is also viscous heating at
large-scales due to pressure anisotropy ($q^+_V$; see
eq. [\ref{eq:pressure_evolution}]).  This {\it is} captured by our
simulations---including the correct ratio of electron to ion heating---and
represents direct conversion of gravitational potential energy
into heat at large scales (see \S2). Figure \ref{fig:fig1} shows a
plot comparing the change in the total plasma energy with the work
done by the boundaries in a typical shearing box simulation.  Energy
is conserved to within $3 \times 10^{-3}$ over a disk rotation period,
which is adequate for our purposes.

\subsection{Stresses and heating rates}

Tables \ref{tab:tab1} \& \ref{tab:tab2} summarize the properties of a
number of simulations with the ion to electron temperature ratio
initialized at values from $1-10^4$ after the MRI saturates at 5
orbits.  We find that the properties of saturated MRI turbulence in a
collisionless plasma are relatively insensitive (to within a factor of
few) to the conductivity (parameterized by $k_L$) or numerical
resolution.  In particular, angular momentum transport quantities like
the anisotropic stress ($\alpha_A$, where $\alpha$ is the stress
divided by the initial pressure), Maxwell stress ($\alpha_M$), and
Reynolds stress ($\alpha_R$), and the associated viscous ($q_V^{+}$),
``numerical resistive" ($q^+_{NR}$), and ``numerical viscous"
($q^+_{NV}$) heating rates do not depend sensitively on $k_L$ or
resolution.  Instead, the most important physical effect in the
evolution of the MRI in a collisionless plasma is pitch angle
scattering by kinetic instabilities (\S 2.2.1), which determines the
magnitude of the anisotropic stress and thus the magnitude of the
viscous heating.  For the pitch angle scattering thresholds described
in \S 2.2.1 we find that anisotropic stress and the associated viscous
heating are the dominant terms in the angular momentum and internal
energy equations.  In addition, viscous heating is dominated by direct
heating via the large-scale shear (eq. [\ref{eq:visc}]; $q^+_{V1}$);
heating due to the turbulent fluctuations (eq. [\ref{eq:fluctuating}];
$q^+_{V2}$) is significantly smaller (see Tables
\ref{tab:tab1} \& \ref{tab:tab2}).

\begin{figure}
\plottwo{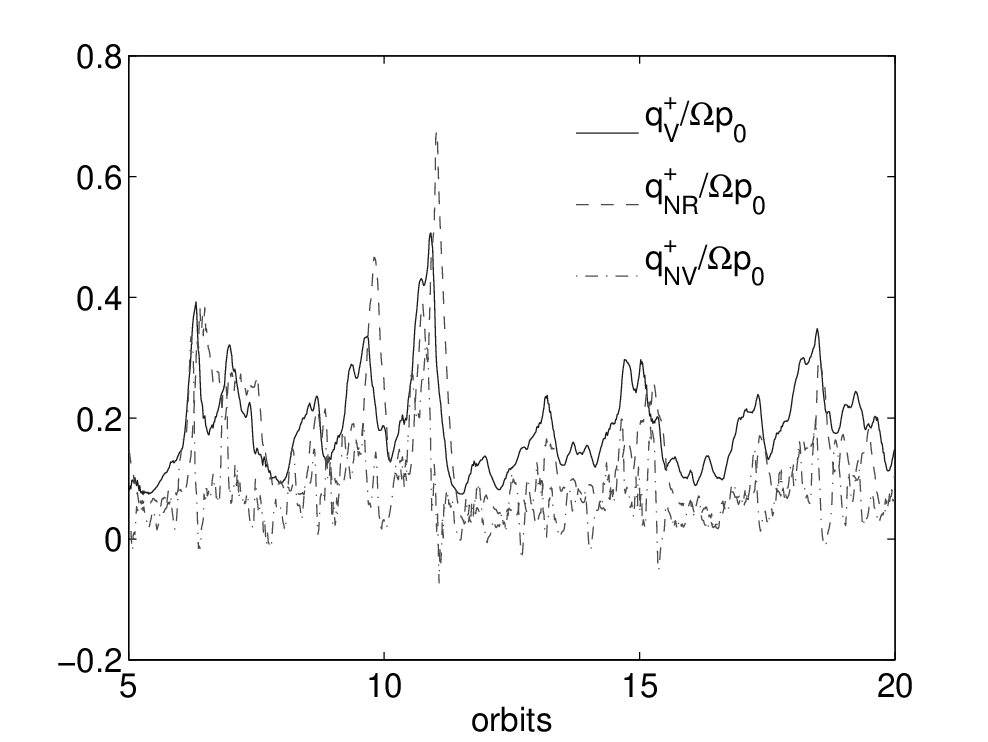}{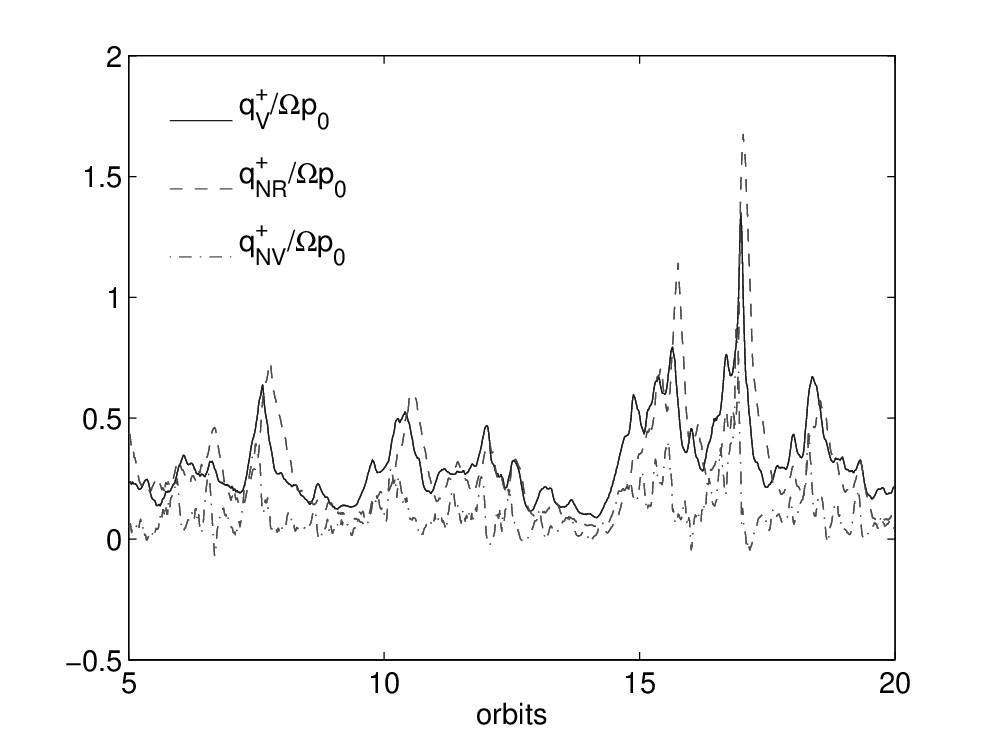}
\caption{Different contributions to plasma heating in shearing box
simulations of the MRI for initially cold electrons ($T_i/T_e = 10^4$
at 5 orbits). {\it Left:} resolution of $27 \times 59 \times 27$
resolution with $k_L=0.5/\delta z$, {\it Right:} resolution of $54
\times 118 \times 54$ with $k_L=0.25/\delta z$. $q_V^+$ is the
heating due to anisotropic viscosity at large scales (electron + ion,
but primarily ion because $T_i \gg T_e$); $q_{NR}^+$, the numerical
loss of magnetic energy; $q_{NV}^+$, the numerical loss of kinetic
energy.  Viscous heating ($q_V^+$) accounts for $\sim 50\%$ of the
heating in our simulations.
\label{fig:fig2}}
\end{figure}

Figure \ref{fig:fig2} shows the volume average of different heating
rates as a function of time in our shearing box simulations with
$T_i/T_e = 10^4$ at 5 orbits, for $N_x \times N_y \times N_z = 27
\times 59 \times 27$ and $k_L=0.5/\delta z$ (left), and $N_x \times N_y
\times N_z = 54 \times 118 \times 54$ and $k_L=0.25/\delta z$ (right).
These combinations of resolution and $k_L$ correspond to the same
conductivity in the two different simulations. Both the high and low
resolution simulations show statistically similar heating results.
Direct viscous heating is slightly larger than numerical resistivity
and viscosity, implying that $\sim 50\%$ of the gravitational
potential energy of the accretion flow is directly converted into heat
at large scales (see also Table \ref{tab:tab1}).  Figure
\ref{fig:fig2} also shows that the different heating rates are
correlated and fluctuate together in time. Although the viscous
heating rate varies as a function of time, we find that the ratio of
the ion to the electron viscous heating does not show as large
statistical variations, i.e., the ion and electron heating rates
increase/decrease together in tandem.\footnote{This is because the ion
and electron pressure anisotropies are approximately equal to the ion
cyclotron (for ions) and electron whistler (for electrons) thresholds,
which have a similar dependence on $\beta$ (see
eq. [\ref{eq:heating_kinetic}]).}

\begin{figure}
\plotone{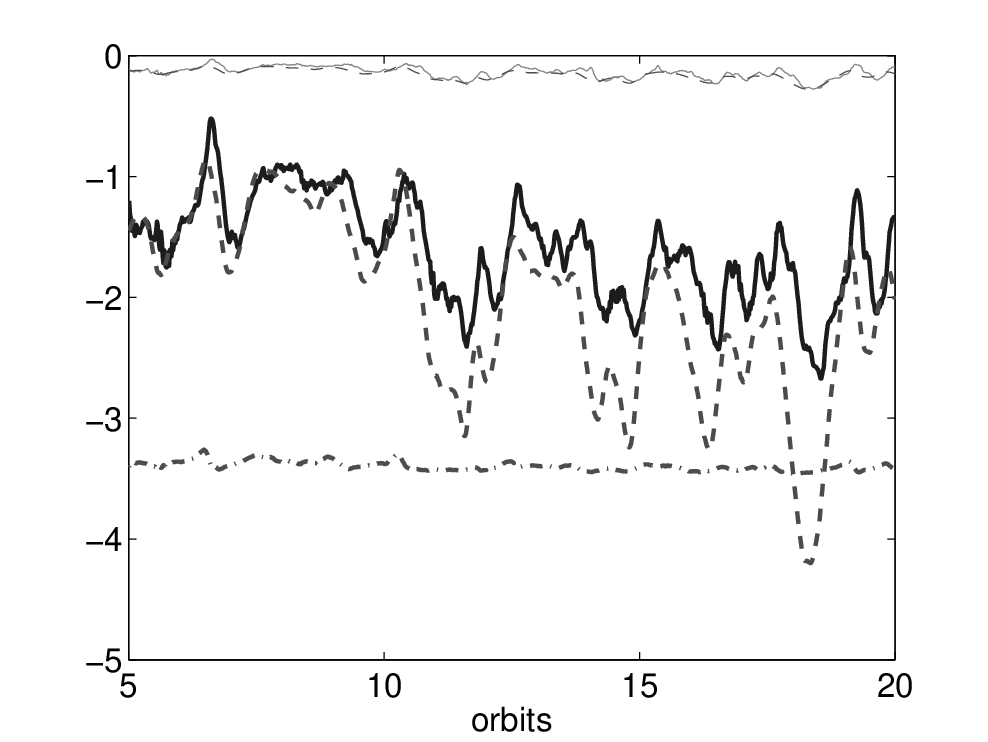}
\caption{Volume averaged pressure anisotropy ($4\pi\Delta p/B^2$; $\Delta
p=p_\Par-p_\Perp$) for electrons (thin solid line) and ions (thick
solid line), compared to the anisotropy thresholds for the ion
cyclotron (thick dashed line), mirror (thick dot-dashed line), and
electron whistler (thin dashed line) instabilities.  The ion to
electron temperature ratio is set to $10$ after the initial saturation
of the MRI at 5 orbits.
\label{fig:fig3}}
\end{figure}

Figure \ref{fig:fig3} shows the volume averaged pressure anisotropy
for electrons and ions with $T_i/T_e$ set to 10 at the end of 5
orbits; the mirror, ion cyclotron, and electron whistler thresholds
are also shown. The ion pressure anisotropy is larger than that of the
electrons for two reasons: first, the electron pressure anisotropy
threshold is smaller than that of the ions ($S_e \sim 0.4 S_i$), and
second, the electron $\beta$ is smaller by a factor of 10.
Figure \ref{fig:fig3} also shows that the electron pressure anisotropy
is relatively close to the electron whistler instability threshold and
that the ion pressure anisotropy roughly tracks the ion cyclotron
threshold.

In a collisionless plasma, the MRI grows faster for an initial
magnetic field configuration with $B_\phi=B_z$ than for $B_\phi = 0$
(Quataert, Dorland, \& Hammett 2002; Sharma, Hammett, \& Quataert 2003).  
However, the nonlinear saturated state is
similar to that of the pure vertical field case described here
\citep[][]{sha06b}.  In particular, the anisotropic stress is again
comparable to the Maxwell stress and viscous heating accounts for a
significant fraction of the total heating of the plasma.

\begin{figure}
\plotone{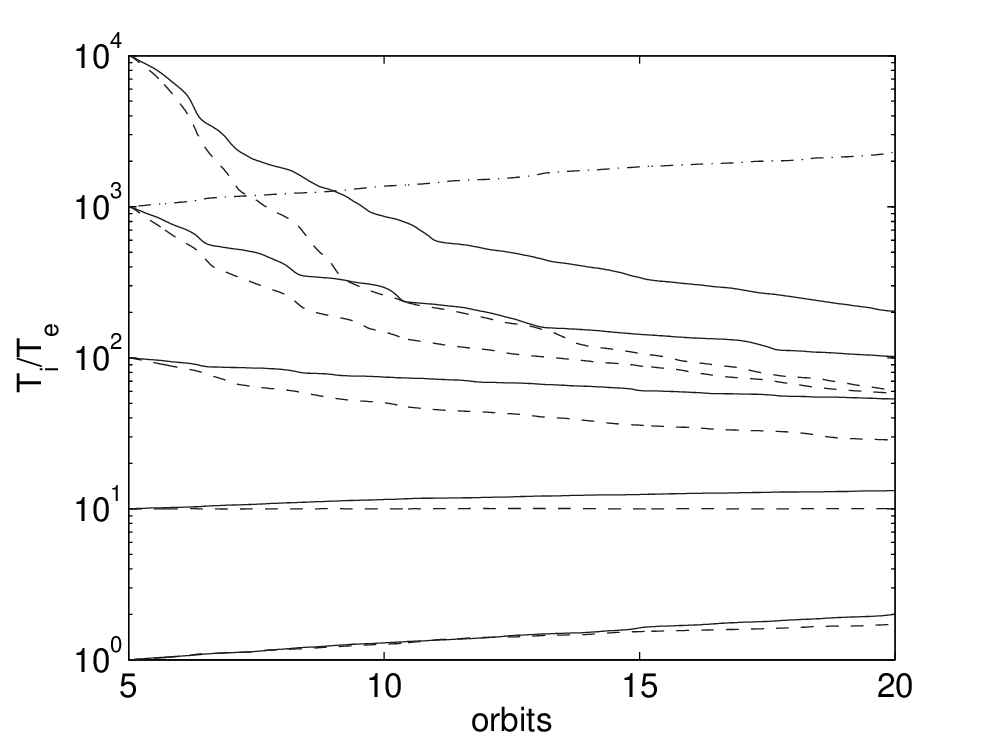}
\caption{The ratio of the volume averaged ion and electron
temperatures as a function of time for models where the ion and
electron temperature ratios are set to $10^4$, $10^3$, $10^2$, 10, and
1 at the end of 5 orbits (top to bottom). Solid lines are for
$k_L=0.5/\delta z$ and dashed lines are for $k_L=\infty$ (the CGL
limit). The electron heating is somewhat larger in the CGL limit (see
also Fig. [\ref{fig:fig5}]). Also shown (dot dashed line) is a
calculation with $T_i/T_e=10^3$ at 5 orbits for $k_L=0.5/\delta z$ in
which viscous heating of electrons is artificially turned off by
setting $q^+_{V,e} = 0$. The electron temperature remains nearly
constant but the ions are heated by viscosity, resulting in an
increase in $T_i/T_e$ with time. This calculation demonstrates that our
simulations can readily sustain large temperature differences between
electrons and ions.
\label{fig:fig4}}
\end{figure}

To quantify the relative heating of ions and electrons as a function
of $T_i/T_e$, we initialize simulations in the saturated turbulent
state (after 5 orbits) with $T_i/T_e = 10^4, 10^3, 10^2, 10,$ \& $1$.
Figure \ref{fig:fig4} shows the ratio of the volume averaged ion and
electron temperatures for $k_L=0.5/\delta z$, and in the CGL limit
($q_\perp=q_\Par=0$). Initially cold electrons become heated
significantly by viscous heating as discussed in \S2.2.  Based on
extrapolating the result of Figure \ref{fig:fig4} to even later times,
we estimate that the late-time value of $T_i/T_e$ due to viscous
heating alone is $\approx 10-30$, in reasonable agreement with the
analytic estimate in \S 2.2.

\begin{figure}
\plotone{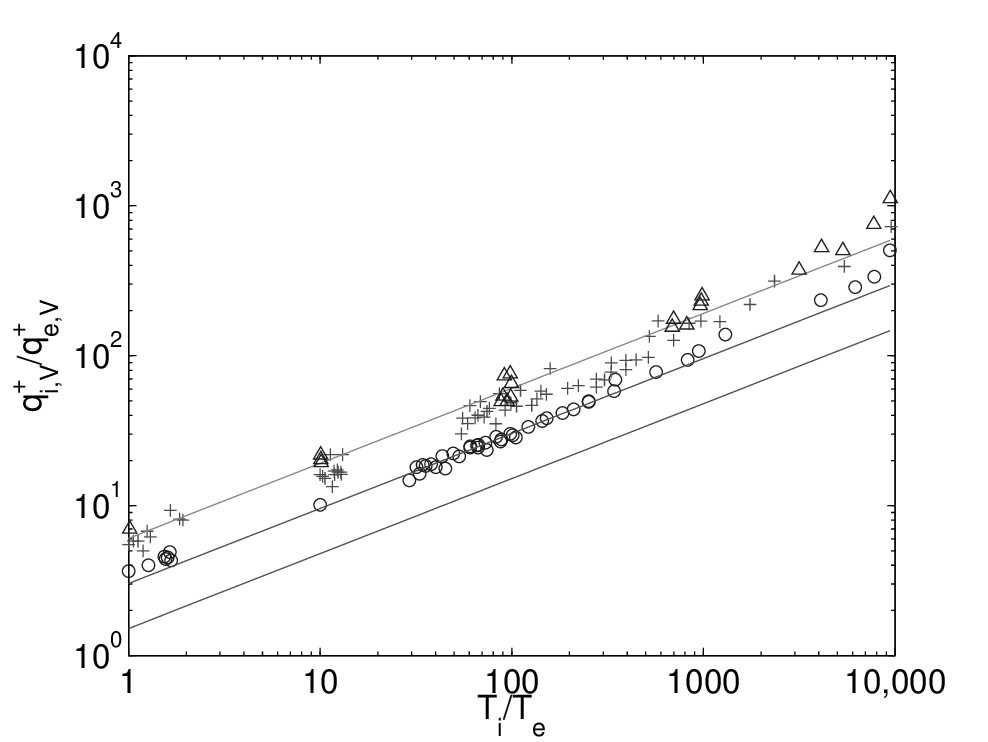}
\caption{The ion to electron heating ratio as a function of the
 temperature ratio for: $k_L=0.5/\delta z$, resolution $27 \times 59
 \times 27$ (+); $k_L=0.25/\delta z$, resolution $54 \times 118 \times
 54$ ($\triangle$); and the CGL limit, resolution $27 \times 59 \times
 27$ (o). The ion to electron heating ratio is slightly larger (less
 than a factor of 2) with thermal conduction. The results with double
 the resolution are comparable to the standard resolution showing that
 the results are reasonably converged. The three solid lines
 correspond to $q^+_{i}/q^+_{e}=$ 6 $(T_i/T_e)^{1/2}$, 3
 $(T_i/T_e)^{1/2}$, and 1.5 $(T_i/T_e)^{1/2}$ (from top to
 bottom). These estimates of the electron heating rate, based on our
 analytic prediction from \S 2.2, approximate the numerical
 simulations reasonably well; they are used to calculate the radiative
 efficiency in \S4 and Figure \ref{fig:fig6}.  For this plot, the
 heating and temperature ratios are averaged over 0.1 orbits in a
 number of simulations (so that $T_i/T_e$ is roughly constant over the
 averaging).
\label{fig:fig5}}
\end{figure}

Figure \ref{fig:fig5} shows the ratio of the volume averaged ion and
electron heating rates as a function of $T_i/T_e$ for the same
calculations as in Figure \ref{fig:fig4} (`o' for CGL, `$+$' for
$k_L=0.5/\delta z$, and `$\triangle$' for a higher resolution
simulations with $k_L=0.25/\delta z$).
This plot was made by averaging the heating and $T_i/T_e$ over 0.1
orbits in a number of different simulations (so that the temperature ratio is
fairly constant over the time of averaging).  The middle solid line in
Figure \ref{fig:fig5} shows $q^+_{V,i}/q^+_{V,e} = 3(T_i/T_e)^{1/2}$,
which is roughly the analytic prediction from \S2.2
(eq. [\ref{eq:heating_kinetic}]) assuming that the ion and electron
pressure anisotropies are set by the ion cyclotron and whistler
instabilities, respectively.  The agreement between the analytic
prediction and the numerical results is particularly good in the
absence of conduction: the heating ratio $q^+_{V,i}/q^+_{V,e}$ is
slightly larger (less than a factor of two) with conduction than
without it.  Figure \ref{fig:fig5} also shows that the electron to
ion heating ratio does not depend sensitively on the resolution of the
simulation.

\section{Implications}

To quantify the importance of viscous heating for observations of
accreting black holes, we use a heating prescription motivated by our
analytic and numerical results in one-dimensional models of RIAFs
(based on \citealt{qua99}).  From equation (\ref{eq:heating_kinetic})
and Figure \ref{fig:fig5},
we approximate the fraction of the viscous energy that heats the
electrons as $\delta \equiv q_e^+/q_i^+ \approx 0.33 (T_e/T_i)^{1/2}$.
This approximation is consistent with our numerical simulations in
the CGL limit; $\delta$ is slightly smaller if thermal conduction is included
(Fig. \ref{fig:fig5}).
Our one-dimensional calculations include electron
cooling by free-free emission, synchrotron radiation, and Inverse
Compton emission.  The electron temperature, the spectrum of
radiation, and the radiative efficiency are calculated
self-consistently given the assumed electron heating rate and the
density, radial velocity, etc. from a one-dimensional dynamical model.
For more details, see \citet{qua99} and references therein.

Figure \ref{fig:fig6} shows the resulting radiative efficiency of RIAF
models as a function of accretion rate; we assumed $\alpha = 0.1$ and
$\beta = 10$ in these calculations but the results are not that
sensitive to reasonable variations in these parameters.  Three choices
of $\delta$ ($\delta = 0.66 [T_e/T_i]^{1/2}$, $\delta = 0.33
[T_e/T_i]^{1/2}$, and $\delta = 0.17 [T_e/T_i]^{1/2}$ ) are shown,
given the uncertainty in the exact magnitude of the viscous heating
(e.g., the relative contribution of viscous heating and other heating
mechanisms, the effects of thermal conduction, and the uncertainty in
the pressure anisotropy thresholds).
For concreteness, the calculations shown in Figure \ref{fig:fig6}
assume that the accretion rate is constant with radius.  Since most of
the radiation is produced at small radii, the radiative efficiency
only depends on the accretion rate at $\sim 10$ Schwarzschild radii
and so the x-axis in Figure \ref{fig:fig6} can be interpreted as such.
Figure \ref{fig:fig6} demonstrates that with analytically and
numerically motivated viscous heating rates, the radiative efficiency
is $\gtrsim 0.5 \%$ for $\dot M \gtrsim 10^{-4} \, \dot M_{\rm Edd}$
for all of our electron heating models considered.
At very low $\dot M$, the efficiency decreases in all models because
the electron cooling time (primarily by synchrotron radiation) becomes
longer than the inflow time in the accretion flow. Since our calculations
in Figure \ref{fig:fig6} do not account for the $\sim 50 \%$ of the
gravitational potential energy lost via grid-scale averaging in our shearing
box simulations (see \S 3), the resulting radiative efficiencies are likely
a lower limit to the true radiative efficiency.

\subsection{Electron collisionality}

One possible caveat to the application of our fully collisionless
results to RIAF models is that if the electron Coulomb collision
frequency $\nu_e$ is significantly larger than the rate at which
pressure anisotropy is created by turbulence, then Coulomb collisions
will suppress the electron pressure anisotropy faster than kinetic
instabilities.\footnote{Ion isotropization is much slower than
electron isotropization and so we do not need to consider a similar
effect for the ions.}  As a result, electron heating by anisotropic
pressure will be negligible.  We thus estimate an upper limit on the
mass accretion rate above which collisions will be able to wipe out
the pressure anisotropy.  This calculation is
analogous to standard estimates in the literature of the critical mass
accretion rate $\sim \alpha^2 \dot M_{\rm Edd}$ above which it is not
possible to maintain a two-temperature RIAF because Coulomb collisions
thermally couple the electrons and protons on an inflow time
\citep[][]{ree82}.  Because electron pitch angle scattering is $\sim
m_p/m_e \sim 10^3$ faster than electron-proton energy exchange, the
critical accretion rate above which the electrons are isotropized
might be expected to be $\sim 10^{-3} \alpha^2 \dot M_{\rm Edd}$.
This result is not correct, however, for two reasons: 1.
Pressure anisotropy is created on a timescale even shorter than an eddy
turnover time due to the fluctuating magnetic field,
rather than on the inflow time (see eq. [\ref{eq:aniso_rate}] discussed below).
2. The electron Coulomb collision frequency depends strongly on
electron temperature, which is itself a strong function of accretion
rate.

The average pressure anisotropy can be estimated from equation
(\ref{eq:pressure_anisotropy}). The second term vanishes after spatial
averaging.  Assuming that fluctuations are incompressible, and
neglecting the heat fluxes, the dominant terms are \be
\label{eq:aniso_rate1}
\frac{\partial}{\partial t} \Delta p_e \sim -3 p_e {\bf
  \hat{b}\hat{b}:\nabla V} - \nu_{eff,e} \Delta p_e - \nu_e \Delta p_e
  \ee where we have added a separate term for the rate $\nu_e$ of
  pitch angle scattering by Coulomb collisions to clearly distinguish
  between the effects of Coulomb collisions and microinstabilities
  ($\nu_{eff,e}$).  If $\nu_e$ is negligible, then steady state occurs
  when the pitch angle scattering caused by microinstabilities, at the
  rate $\nu_{eff,e}$ balances the first term on the RHS of equation
  (\ref{eq:aniso_rate1}). The scattering by microinstabilities occurs
  very rapidly (on the electron cyclotron timescales) if the pressure
  anisotropy is larger than the threshold to excite the instabilities,
  so in practice $\nu_{eff,e}$ will always be just large enough to
  keep $\Delta p_e$ near the threshold. Thus balancing the first and
  second terms on the RHS gives \be
\label{eq:aniso_rate}
\nu_{eff,e} \sim 3 \frac{p_e}{\Delta p_e} {\bf \hat{b}\hat{b} : \nabla V} \sim 35 \Omega
\beta^{1/2} \left ( \frac{T_e}{T_i} \right )^{1/2},
\ee
using $\alpha_e$ and $S_e$ from \S2.2.1.
If the collisional pitch angle scattering rate exceeds that estimated
in equation (\ref{eq:aniso_rate}), then electron viscous heating will be
reduced relative to the pure collisionless calculations presented in \S 2 \& 3.
Figure \ref{fig:fig7} shows this explicitly by plotting the average electron
heating rate for shearing-box simulations that include an additional Coulomb collision
rate $\nu_e$, in addition to the subgrid models for kinetic
instabilities.  The numerical results in Figure \ref{fig:fig7} show that above a critical
$\nu_e$,  electron viscous heating is substantially reduced; this critical
$\nu_e$ depends on $T_e/T_i$, in good agreement with the analytic estimate
in equation (\ref{eq:aniso_rate}).

To apply the above results to RIAF models, we note that for the accretion rates of
interest, the electrons are only marginally relativistic (see
Fig. \ref{fig:fig8} discussed below) and so we use the Coulomb
collision frequency for non-relativistic electrons.
Using the ADAF solution of \citet{nar95} to estimate the density and
ion temperature in the accretion flow, we find that electron Coulomb
collisions are negligible for the isotropization of the electron
pressure for accretion rates satisfying
\be
\label{eq:Mdot_crit}
{\dot M \over \dot M_{\rm Edd}} \ll 0.06 \left (\frac{\alpha}{0.1}
\right ) \left ( \frac{\beta}{10} \right )^{1/2} \left ( \frac{r}{10}
\right )^{1/2} \theta_e^2, \ee where $r$ is the distance from the
central object in units of the Schwarzschild radius and
$\theta_e=kT_e/m_ec^2$ is the electron temperature normalized to the
electron rest mass energy.

Figure \ref{fig:fig8} shows the electron temperature as a function of
radius for our one-dimensional RIAF models with $\delta = 0.33
(T_e/T_i)^{1/2}$, for a variety of accretion rates (solid lines).  The
protons are at roughly the virial temperature for all $\dot M$ since
they do not cool significantly.  By contrast, the electron temperature
depends strongly on $\dot M$, with the temperature decreasing at small
radii for increasing $\dot M$.  For $\dot M \lesssim 10^{-5} \dot
M_{\rm Edd}$, however, the electron temperature is relatively
independent of $\dot M$ because the electron cooling time becomes long
compared to the inflow time.  The dotted line shows the electron
temperature for low $\dot M$ for $\delta = 0.66(T_e/T_i)^{1/2}$; the
temperature is a factor of $\sim 2$ larger at small radii in this case
because of the additional heating.

Figure \ref{fig:fig8} shows that $\theta_e \sim 1$ at $\dot M \sim
10^{-2} \dot M_{\rm Edd}$; at this high $\dot M$, the electron
temperature is relatively independent of our assumptions about the
electron heating. This is because the electrons are efficiently cooled
by synchrotron radiation and Inverse Compton emission, and become
marginally relativistic.
Using $\theta_e \sim 1$ in equation (\ref{eq:Mdot_crit}), we conclude
that electron Coulomb collisions are unimportant for isotropizing the
electrons for $\dot M \ll 0.06 \dot M_{\rm Edd}$.  Since the electron
temperature increases with decreasing $\dot M$ (Fig. \ref{fig:fig8}),
Coulomb collisions rapidly become unimportant for isotropizing the
electrons below $\sim 0.06 \dot M_{\rm Edd}$.  We thus conclude that
the collisionless calculations of the radiative efficiency in Figure
\ref{fig:fig6} should be applicable for essentially all of the
accretion rates considered.

\section{Discussion}

We have shown that viscosity arising from anisotropic pressure is a
significant source of heating in hot accretion flows (RIAFs).  In
shearing box simulations of the MRI in collisionless plasmas, viscous
heating is comparable to the rate at which energy is lost by
grid-scale numerical averaging.  It thus accounts for $\sim 50\%$ of
the gravitational potential energy released by the inflowing matter.
In a real accretion disk, the fate of the remaining $\sim 50\%$ of the
energy that is dissipated at small scales is unclear.  Although
viscosity along the field lines due to pressure anisotropy can damp
parallel motions, it cannot dissipate energy in motions perpendicular
to the field lines (e.g., Alfv\'en modes).  This energy is presumably
dissipated through
collisionless damping of fluctuations in a turbulent cascade at the
ion Larmor radius \citep[][]{qua98,how06}.

In contrast to the kinetic and magnetic energy lost to grid-scale
averaging, the physics of viscous heating at large scales is
well-captured by MRI simulations and can be accurately approximated
using a simple analytic expression that depends primarily on the
average pressure anisotropy in the plasma (eq.
[\ref{eq:heating_kinetic}]).  In turn, the magnitude of the pressure
anisotropy is set by pitch angle scattering due to small-scale kinetic
instabilities (see \S 2.2.1).  We thus conclude that pressure
anisotropy plays an essential role in both the energy (via viscous
heating) and angular momentum (via anisotropic stress) budgets of
RIAFs. This interplay of pressure anisotropy and microinstabilities is
likely to be important in many other weakly collisional plasmas, e.g.,
the solar wind and X-ray clusters \citep[e.g.,][]{sch05}. The only way
out of this conclusion is if there is efficient generation of high
frequency fluctuations that provide pitch angle scattering (and thus
pressure isotropization) at a rate much faster than that of the
kinetic instabilities discussed in \S 2.2.1.  This does not appear to
be true in the solar wind, where large pressure anisotropies are
measured \citep[e.g.,][]{cra99} and where there is {\it in situ}
experimental evidence for pressure isotropization via the firehose
instability \citep[][]{kas02}.

In addition to viscous heating, numerical simulations can in principle
also capture heating by the collisionless (Landau/Barnes) damping of
large-scale turbulent fluctuations.  In particular, the slow
magnetosonic mode is strongly damped in a collisionless plasma even on
large scales and has been proposed as a source of proton heating in
RIAFs \citep[e.g.,][]{qua98,black99}.\footnote{Heating by Alfv\'enic
turbulence is not expected to be important until very small scales of
order the ion Larmor radius, scales that are not resolved by our MRI
simulations.  The fast mode is also strongly damped by collisionless
damping, but is not likely to be as efficiently excited by the weakly
compressible MRI turbulence.} We find little direct evidence for such
heating in our numerical simulations; the part of the anisotropic
pressure heating (eq. [\ref{eq:pressure_evolution}]) proportional to
the background shear $d\Omega/d\ln r$ (eq. [\ref{eq:visc}]) dominates
over the part proportional to the turbulent velocity fluctuations
(eq. [\ref{eq:fluctuating}]) for all of our simulations (see Table
\ref{tab:tab1}). In addition, the relatively weak dependence of the
plasma heating on the Landau damping parameter $k_L$ (see Tables
\ref{tab:tab1} \& \ref{tab:tab2}) suggests that very little of the
turbulent energy is dissipated via collisionless damping at large
scales.  In fact, even in the limit of zero conductivity ($k_L
\rightarrow \infty$) and thus no collisionless damping (the
double-adiabatic limit) we find roughly the same energetics in our
shearing box simulations.  These results imply that heating due to
work done by anisotropic stress dominates collisionless damping on
large scales in RIAFs. Collisionless damping may become important on
smaller scales (that we cannot resolve) where the turbulence can be
better approximated as a superposition of linear modes (on large
scales the fluctuating magnetic field is much larger than the mean
field). Indeed, with higher resolution it may turn out that instead of
being lost to grid-scale averaging, most of the kinetic and magnetic
energy is damped via collisionless damping at intermediate scales.  In
addition, more accurate fully kinetic treatments may be needed to
quantify the heating by collisionless damping on large scales in MRI
simulations.

\begin{figure}
\epsscale{0.9}
\plotone{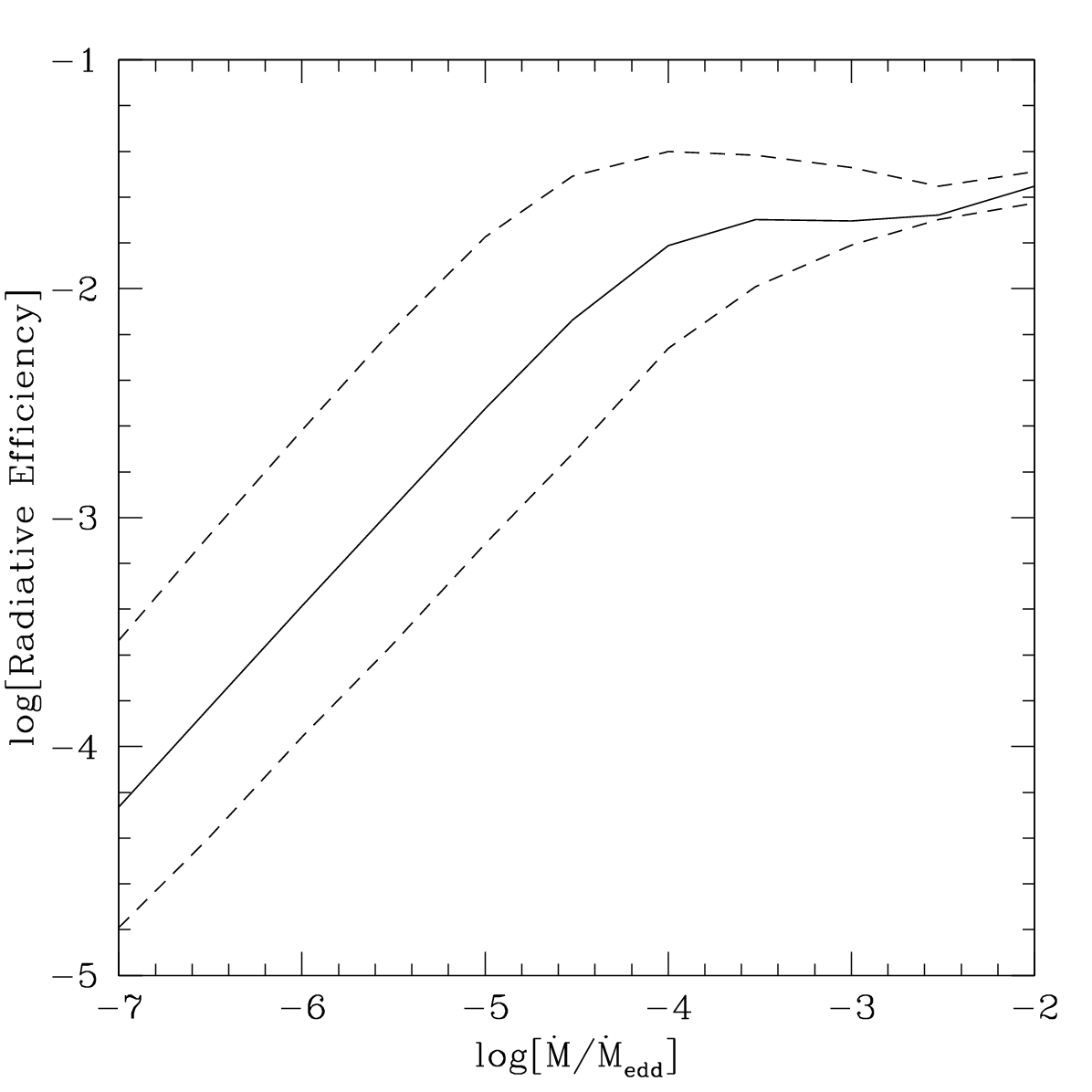}
\caption{Radiative efficiency as a function of mass accretion rate in
one dimensional models of RIAFs (based on the calculations described
in \citealt{qua99}).  These models assume $\alpha = 0.1$ and
$\beta = 10$ in determining the density and magnetic field strength in
the accretion flow.  Since most of the radiation is produced close to
the black hole, $\dot M$ can be interpreted as the accretion rate in
the inner $\sim 10 R_S$. The solid line corresponds to $\delta
\equiv q^+_e/q^+_i=0.33 (T_e/T_i)^{1/2}$, and the dashed lines to $\delta =0.66
(T_e/T_i)^{1/2}$ (upper) and $\delta=0.17(T_e/T_i)^{1/2}$ (lower).
At high $\dot M$, Coulomb collisions transfer energy from the ions to
the electrons and so the radiative efficiency becomes relatively
independent of the assumptions regarding $\delta$.  At very low $\dot
M$, the electron cooling time is longer than the inflow time and the
radiative efficiency decreases in all models.\label{fig:fig6}}
\end{figure}

\begin{figure}
\plotone{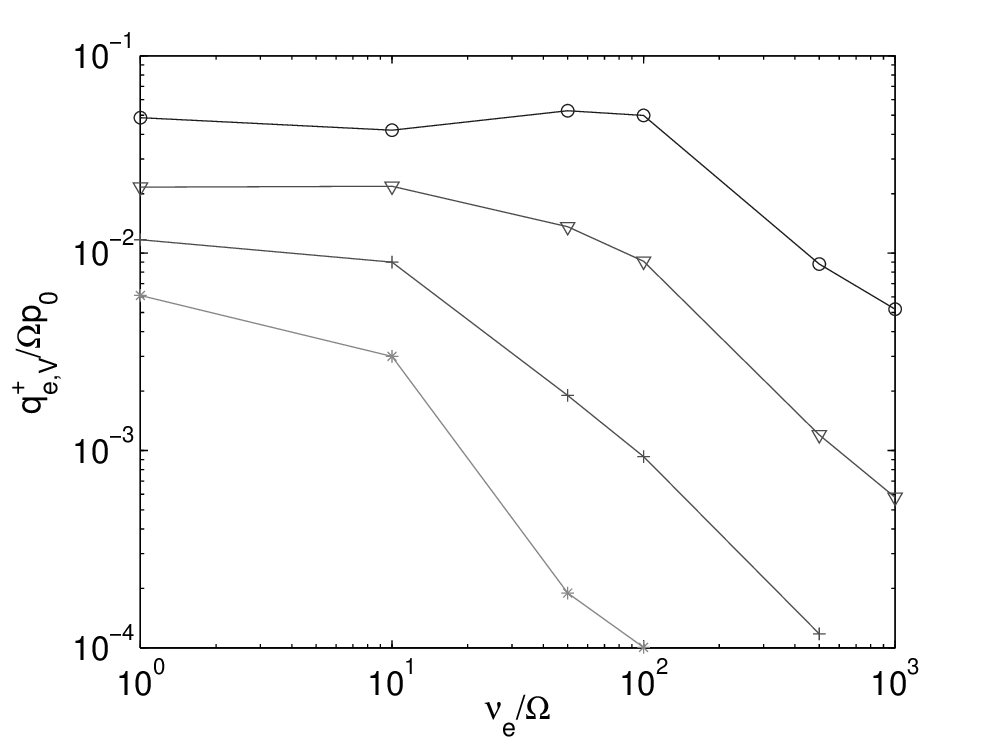}
\caption{Electron heating rate (volume averaged from 5 to 20 orbits)
due to anisotropic viscosity as a function of the electron
collisionality for simulations with no heat conduction ($k_L
\rightarrow \infty$). The ion to electron temperature ratio is
initialized to $10^{3}$ ($*$),  $10^{2}$ (+), 10 ($\triangledown$), \&
$1$ (o) after 5 orbits.  Viscous heating of electrons is suppressed at
high electron collisionality (high $\nu_e$); the critical value of
$\nu_e/\Omega$ in the numerical calculations is in reasonable
agreement with the analytic estimate in equation
(\ref{eq:aniso_rate}).\label{fig:fig7}}
\end{figure}

Our results on viscous heating imply radiative efficiencies of
$\gtrsim 0.5 \%$ for $\dot M \gtrsim 10^{-4} \dot M_{Edd}$
(Fig. \ref{fig:fig6}).  With such significant radiative efficiencies,
the low luminosities of many accreting black holes can only be
understood if most of the mass supplied to the flow at large radii
does not reach the central object.  Our results are thus consistent
with global numerical simulations of magnetized RIAFs, which show that
only a small fraction of mass is actually accreted (e.g.,
\citealt{sto01,haw01}).

\begin{figure}
\epsscale{0.9}
\plotone{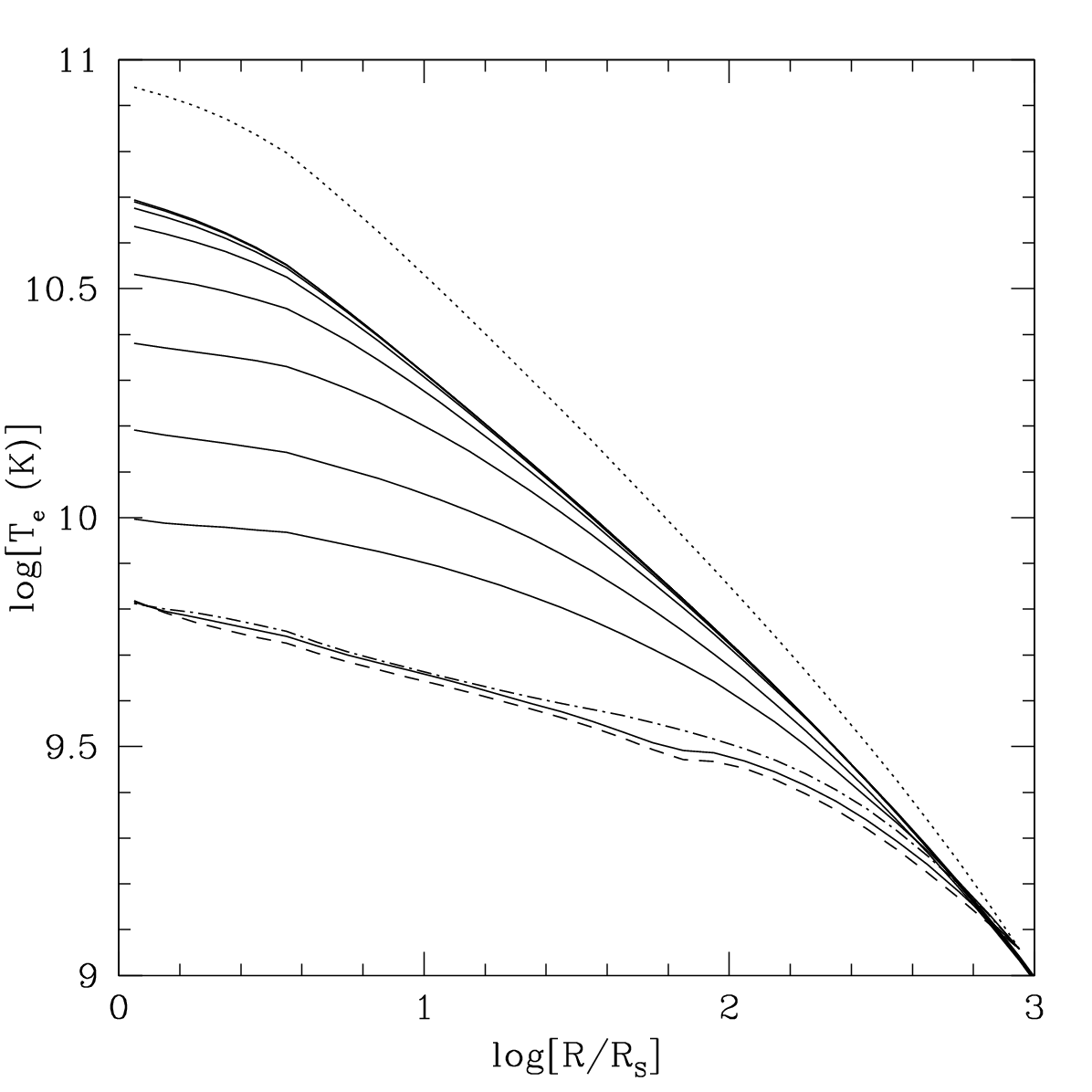}
\caption{Electron temperature as a function of radius for our
one-dimensional RIAF models with $\alpha = 0.1$, $\beta = 10$, and
$\delta = 0.33 (T_e/T_i)^{1/2}$.  The solid lines correspond to $\log
[\dot{M}/\dot M_{\rm Edd}] = $ -2, -2.5, -3, -3.5, -4, -4.5, -5, -5.5,
-6 (from bottom to top). The electron temperature is independent of
$\dot M$ for even lower accretion rates. The dotted line shows the
electron temperature for $\delta = 0.66 (T_e/T_i)^{1/2}$ and $\log
[\dot{M}/\dot M_{\rm Edd}] = -6$.  To demonstrate that at high $\dot
M$ the electron temperature is relatively independent of our model
assumptions, the dashed line shows $T_e(r)$ for $\dot M = 10^{-2} \dot
M_{\rm Edd}$ and $\delta = 10^{-3}$, while the dot-dashed line is for
$\dot M = 10^{-2} \dot M_{\rm Edd}$ and $\delta = 0.66
(T_e/T_i)^{1/2}$.
\label{fig:fig8}}
\end{figure}

To give a concrete example of the implications of our results, the
Bondi accretion rate for Sgr A* in the Galactic Center is estimated to
be $\sim 3 \times 10^{-6} - 10^{-5} M_\odot \, {\rm yr^{-1}}$
\citep[e.g.,][]{qua04,cua06}, which corresponds to $\dot M/\dot M_{\rm
Edd} \sim 10^{-4}$.  Our results predict a radiative efficiency of
$\approx 0.005-0.03$ for this $\dot M$, with the uncertainty in the
radiative efficiency arising from uncertainties in the exact rate of
electron viscous heating.  If gas were to accrete at the Bondi rate, a
radiative efficiency of $0.01$ implies a luminosity $\sim 10^3$ times
larger than what is observed from Sgr A*, implying that the accretion
rate must be well below the Bondi rate.  Requiring $L \approx 10^{36}$
ergs s$^{-1}$ as is observed, our predicted radiative efficiencies
(Fig. \ref{fig:fig6}) imply an accretion rate of $\dot M \approx
10^{-7}-10^{-8} \, {\rm M_\odot \, yr^{-1}}$ where the range accounts
for the range of radiative efficiencies in Figure \ref{fig:fig6}.
Measurements of the rotation measure from Sgr A* are consistent with
such a low accretion rate \citep[][]{mar07}. In addition, VLBI
interferometry of Sgr A* finds a size of $\approx 15 R_S$ at 100 GHz
\citep[][]{bow04,bow06}.  The corresponding brightness temperature is
$\approx 3 \times 10^{10}$ K.  At low $\dot M$, our models for the
electron temperature at $\approx 15 R_S$ are in reasonable agreement
with this result (Fig. \ref{fig:fig8}).

\acknowledgements
PS and EQ were supported in part by NASA grant
NNG06GI68G, an Alfred P. Sloan Fellowship, and the David and Lucile Packard Foundation.
PS was partly supported by DOE award DE-FC02-06ER41453 to Jon Arons.
GWH is supported by NASA Grant NNH06AD01I and by U.S. DOE contract No. DE-AC02-76CH03073.
JS was supported by DOE grant DE-FG52-06NA26217.
Most of the computing resources were provided by the Princeton Plasma Physics Laboratory
Scientific Computing Cluster. This research also used resources of the National Energy
Research Scientific Computing Center, which is supported by the Office of Science of the
U.S. Department of Energy under Contract No. DE-AC02-05CH11231.

\clearpage

\begin{deluxetable}{ccccccccc}
\rotate
\tablecaption{Heating Diagnostics for Simulations with initially Cold Electrons ($T_i/T_e = 10^4$ at 5 orbits)
\label{tab:tab1}}
\tablewidth{0pt}
\tablehead{
\colhead{$k_L$} & \colhead{$N_x \times N_y \times N_z$} & \colhead{$\alpha_A$\tablenotemark{a}} &
\colhead{$\alpha_M$\tablenotemark{b}} & $\alpha_R$\tablenotemark{c} & $q_{V1}^+/\Omega p_0$\tablenotemark{d} &
\colhead{$q_{V2}^+/\Omega p_0$\tablenotemark{e}} & \colhead{$q_{NR}^+/\Omega p_0$\tablenotemark{f}} &
\colhead{$q_{NV}^+/\Omega p_0$\tablenotemark{g}}
}
\startdata
$0.5/\delta z$ & $27 \times 59 \times 27$ & 0.12 & 0.097 & 0.048  & 0.18 & 0.049 & 0.14 & 0.074  \\
$0.5/\delta z$ & $54 \times 118 \times 54$ & 0.21 & 0.26 & 0.083 & 0.31  & 0.06 & 0.38 & 0.13  \\
$0.25/\delta z$ & $54 \times 118 \times 54$ & 0.18 & 0.21 & 0.073 & 0.27 & 0.043 & 0.3 & 0.11 \\
$0.125/\delta z$ & $27 \times 59 \times 27$ & 0.16 & 0.21  & 0.084  & 0.25  & 0.04 & 0.32 & 0.13  \\
$0.125/\delta z$ & $54 \times 118 \times 54$ & 0.15 & 0.20 & 0.060 & 0.23 & 0.049 & 0.31 & 0.094  \\
$2/\delta z$ & $27 \times 59 \times 27$ & 0.16 & 0.15 & 0.073  & 0.23 & 0.016 & 0.22 & 0.11  \\
$2/\delta z$ & $54 \times 118 \times 54$ & 0.19 & 0.22 & 0.071 & 0.28  & 0.039 & 0.30 & 0.11  \\
$\infty$ & $27 \times 59 \times 27$ & 0.19 & 0.20  & 0.086  & 0.27  & 0.013 & 0.25 & 0.13  \\
$\infty$ & $54 \times 118 \times 54$ & 0.18 & 0.22 & 0.067 & 0.27 & 0.026 & 0.27 & 0.1 \\
\enddata
\tablenotetext{a} {$\alpha_A=\langle \langle (p_\Par-p_\Perp)
(B_xB_y/B^2)/ p_0 \rangle \rangle$, $\langle \langle \rangle \rangle$
denotes time and volume average in the turbulent state (from 5 to 20 orbits);
$p_0$ is the initial pressure.} \tablenotetext{b}
{$\alpha_M=\langle \langle -B_xB_y/4\pi p_0 \rangle \rangle$}
\tablenotetext{c} {$\alpha_R=\langle \langle \rho V_x \delta V_y/p_0
\rangle \rangle$, where $\delta V_y = V_y + (3/2) \Omega x$}
\tablenotetext{d} {$q_{V1}^+=q_{V1,i}+q_{V1,e}$, where $q_{V1,s}$ is
given in eq. [\ref{eq:visc}]} \tablenotetext{e}
{$q_{V2}^+=q_{V2,i}+q_{V2,e}$, where $q_{V2,s}$ is given in
eq. [\ref{eq:fluctuating}]} \tablenotetext{f} {Numerical loss of
magnetic energy at the grid-scale} \tablenotetext{g} {Numerical loss
of kinetic energy at the grid-scale}
\end{deluxetable}

\begin{deluxetable}{ccccccccccc}
\rotate
\tablecaption{Heating Diagnostics for Different $T_i/T_e$ \label{tab:tab2}}
\tablewidth{0pt}
\tablehead{
\colhead{$T_i/T_e$\tablenotemark{a}} & \colhead{$k_L$} & \colhead{$\alpha_A$} &
\colhead{$\alpha_M$} & $\alpha_R$ & $q_{V1,i}^+/\Omega p_0$\tablenotemark{b} &
\colhead{$q_{V2,i}^+/\Omega p_0$\tablenotemark{b}} &
\colhead{$q_{V1,e}^+/\Omega p_0$\tablenotemark{b}} &
\colhead{$q_{V2,e}^+/\Omega p_0$\tablenotemark{b}} & \colhead{$q_{NR}^+/\Omega p_0$} &
\colhead{$q_{NV}^+/\Omega p_0$}
}
\startdata
1000 & $0.5/\delta z$ & 0.16 & 0.17 & 0.077 & 0.16 & 0.0037 & $5.9 \times 10^{-4}$ & $1.6\times 10^{-4}$ & 0.26 & 0.12 \\
1000 & $\infty$ & 0.15 & 0.13 & 0.065 & 0.16 & -0.0014 & $8.2 \times 10^{-4}$ & $2.9 \times 10^{-4}$ & 0.17 & 0.1 \\
100  & $0.5/\delta z$ &0.13 & 0.11 & 0.057 & 0.16 & 0.0039 & 0.003 & $3.1 \times 10^{-4}$ & 0.17 & 0.088 \\
100 & $\infty$ & 0.18 & 0.17 & 0.079 & 0.16 & -0.0024 & 0.0042 & $6.9 \times 10^{-4}$ & 0.23 & 0.12 \\
10   & $0.5/\delta z$ & 0.16 & 0.15 & 0.071 & 0.16 & 0.0027 & 0.01  & $3.7 \times 10^{-4}$ & 0.23 & 0.11 \\
10 & $\infty$ & 0.15 & 0.12 & 0.061 & 0.16 & -0.0055 & 0.014  & 0.0012 & 0.16 & 0.09 \\
1    & $0.5/\delta z$ & 0.17 & 0.13 & 0.07  & 0.16 & -0.0015 & 0.03  & $2.1 \times 10^{-4}$ & 0.2  & 0.11 \\
1   & $\infty$ & 0.18 & 0.13 & 0.068  & 0.16 & -0.018 & 0.035  & 0.0056 & 0.17  & 0.1 \\
\enddata
\tablenotetext{a} {The simulation is restarted after 5 orbits with this initial temperature ratio.
$N_x \times N_y \times N_z = 27 \times 59 \times 27$.
All quantities have the same definitions as in Table \ref{tab:tab1}.}
\tablenotetext{b} {Because the electron and ion temperatures change significantly from 5 to 20 orbits,
the heating rates are only averaged from 5 to 5.1 orbits so that the temperature is roughly equal to
the value it is initialized at (column 1).}
\end{deluxetable}

\clearpage

\end{document}